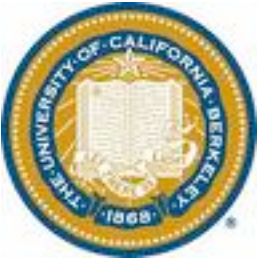

# On the feasibility of multi-polymer, liquid-crystal silica photovoltaics: simulating diodic p-n junctions with ionic gradients

**ARJAN SINGH PUNIANI**

University of California, Berkeley, Center for Theoretical Physics
Department of Chemistry and Chemical Engineering
Research Completed: August 2009

Conventional photovoltaic machinery, including traditional silicone panels, fails to address efficiency problems and is limited in scope in terms of practical applicability. Recent technological advances suggest less metal-specific reliance, but plastic substrates are bound by cost-inefficiency. Photovoltaic paint effectively dissociates from metal dependency and relies on a combination *p-n junction* diode principle/thermoelectric effect to generate electrical energy from solar exposure. Replicating the junction is accomplished via multi-polymer layers of crystalline-silica water-based paint with ionic solution concentration gradient overlap, reconstructing the depletion zone and, in thermal respects, construes the thermoelectric effect via replication of a heavily modified thermocouple. Experimentation revealed the largest gradient (50%-10%) of ionic solution, specifically, sodium-chloride solution, per paint solution liter generated the largest electrical energy yield, with traditional, unadulterated paint yielding none. Maintenance of a functional electro-conductive gradient is achieved with specialized, non-acidic solution, but the lifespan of the charge is virtually instantaneous. The experimentation yielded possible exploits for multi-polymer conduction layer advancement; however, the production of solar-receptive paints replicating the capture and distribution scheme of solar panels, sans the inflexibility and limited application. Prospective applications include electrical car augmentation, solar-receptive buildings, and vast tracts of solar farms given solid initial conditions. Weathering, life expectancy, and storage are paint-specific. The intentions of sculpting the ideal alternative energy are in efforts to propose novel mechanisms curbing fossil fuel dependence and arousing intellectual curiosity and creativity on the many channels converging on innovative environmental feats.

Photovoltaic materials generate an electrical current via the capture of solar energy. Sunlight, traveling in high-energy packets known as photons, interacts with said photo-receptive material and the wafers of semiconductors absorb the particles. Electrons, stripped from this semiconductor by the potency of the photons, navigate through the material freely and unhinged to atoms. The subatomic particles produce electricity with their unadulterated pathways as they flow through the semiconductors. Positive charges are complementarily created, known as "holes," as the electron is excited to a new shell by the quantum jump, and the void is assigned a positive charge.[i]

The conduction layer these free electrons travel to cause the generation of an electrical current to occur. The photoreceptive capabilities stem from the *p-n junction diode* principle—specifically, the proximity of a semiconductor[ii] (p-type) with a doping-capable (impurity-injected purities), poorly insulated semiconductor (n-type). The electromotive force electrons wedge themselves into the material, and the p-type repels holes while n-types repel the electrons, shrinking the depletion zone and voltage potential, until the resistance plunges to miniscule levels and a current is established—all courtesy of forward bias[iii].

For the majority reference, the traditional photovoltaic cell, say, a 4" square single crystal solar cell, which pumps out about .89 watts. Considerable attention mandates importance to the fact that although the current rating or amperage of each cell can vary, a single PV cell, no matter how large or small, will only put out .5 volts (half a volt). Watt output is equal to the product of voltage and amperes, take, by example, .5V * 1.78A = .89 Watts (W).

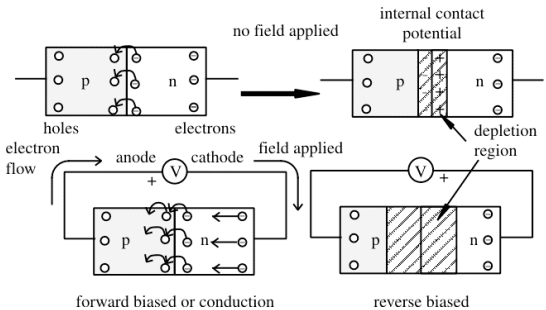

Most panels, for a variety of design reasons, contain between 32 and 36 individual solar cells, with a 36-cell panel emanating more voltage than its less quantified, 32-cell panel. The higher voltage presents utility if one attempted design of panels for a location tending to possess a multitude of cloud cover, as the panel can produce more watts with less sun. My experiment attempts to derive a paint-based substitute for the solidified metals, yet with the retention of the *p-n junction diode* principle.

The thermoelectric effect, a phenomenon coupled with the electrically-conjoined dissimilar metal compounds industrially utilized for temperature readings, can be replicated as a substitute for the *p-n junction diode* principle, and the concept works quite simply: liquefy key photoreceptive metals, such as copper, selenium, or iridium, and attempt to recreate the generated electromotive force.

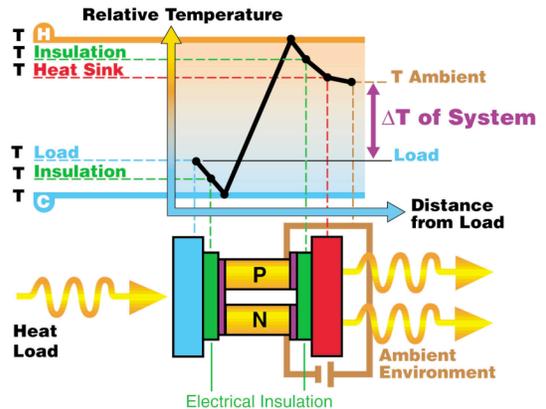
*The thermoelectric effect was slightly modified in response to the depletion zone mechanics.*

The methodologies for producing maximum voltage whilst remaining cost-effective are paramount to experimental procedures, and metals producing the maximum electro-conductive generation have too high of a melting point to feasibly construct a photovoltaic paint capable of retaining the properties in a practical environment. Therefore, it is in best interest to replace these metals with a substitute—a solution capable of generating a current—and such a solution must equivalently dissolve in the paint conglomerate. In all probability, the solution likely to effectively replicate the voltage generation of the Copper-Selenium-Iridium conglomerate is a saline solution.

The simple electrolysis principle is the addition of ions constituting a voltage-viable solution, the most rudimentary of which is the injection of sodium chloride into a distilled water solution. The aforementioned conglomerate mixture fusing these liquids implies the following: a photoreceptive paint capable of application on virtually any surface with the functionality of a conventional solar panel[iv].

What distinguishes this project from contemporary solutions is the cost-efficiency of its implications, as Nanosolar recently developed a material of metal oxide nanowires that can be sprayed as a liquid onto a plastic substrate where it self-assembles into a photovoltaic film. A roll-to-roll process similar to high-speed printing offers a high-volume approach that doesn't require high temperatures or vacuum equipment.[v] Nanosys intends for its solar coatings--based on structures called nanotetrapods—to be sprayed onto roofing tiles, while Konarka broods in the process of developing plastic sheets embedded with titanium dioxide nanocrystals, coated with light-absorbing dyes. The company recently surprised the global, scientific community by acquiring Siemens' organic photovoltaic research activities, and Konarka's recent $18 million third round of funding included the world's first- and fifth-largest energy companies, Electricité de France and ChevronTexaco[vi].

Conventional solar panels, burdened by the inefficiency in solar-electrical output conversion, consumption of lengthy and stretched intervals of time prior to cost-efficiency attainment, limited localization, and evident net-energy loss due to mined materials for construction. Essentially, the paint functions as an electrolyte[vii], as free ions swirl in the solution and act as an electrically conductive medium, and the variance in charge from differing electrolytes with quantized concentrations of ions are able to release differing amounts of charge across the depletion zone until the electrons overpower the insulation and establish a flowing current.

The electrolyte principle is elementary, and frequently natural in the manifestations of acids, bases, or neutralized salts. The solution normally arises when a salt is placed into a solvent, such as distilled water, and the individual atomic components are separated by the force applied upon the solute molecule—in a process known as chemical dissociation—in which the solvent applies said force to effectively hold the ions apart. Salts are compounds that are linked by weak ionic bonds, and will separate into charged ions in the presence of a solvent possessing more powerful covalent bonds. To reiterate, the electrolyte refers to the material that dissolves in a specified solution to grant the said solution a conduction layer to propagate an electrical current.

If this electrolyte solution depends on both a difference of potential and the *p-n junction diode* principle, then no direct, or theoretical, prediction can be made regarding the voltage generation. In fact, the revelation grants a more than appropriate assumption: the photovoltaic paint theory effectively combines previously isolated theories and combines separate physical fields of study to generate voltage. Armed with this knowledge, we extend to the problem statement:

> Conventional solar panels are relatively inefficient in solar-electrical output conversion, consume lengthy and stretched intervals of time prior to cost-efficient attainment, are often limited to confined and specialized locales, and present a net-energy loss when harvesting materials for its construction (mined aluminum ore and semi-conducting materials, such as magnesium and silicone)[viii]. Therefore, in the interest of global concern regarding environmental vitality, a substitute capable of eliminating the problems demands urgent investment. The question derives the sustenance of photovoltaic paint; specifically, does such a novel approach utilizing liquefied electro-receptive materials offer a viable alternative to the conventional solar panels and streamline the hindrances?

Any scientific inquiry beckons for a replicable prosposal:

> Extensive informative research extrapolates a model (primarily derived from the *p-n junction diode* principle and thermoelectric effect) to generate voltage. By fusing a traditional, polycarbonate and water-based painting solution with an ionic solution of varying concentrations of conductivity, let the experimentation be shown of electrical current generation via the energy resonance and temperature gradient of solar power reacting with the prepared photovoltaic paint conglomerate. It is of firm belief that such a substitution of unconventional photovoltaics yields a more cost-effective and efficient proposal than conventional solar arrays.

The purpose of this experimentation is to determine and construct a cost-efficient alternative to conventional photovoltaic panels by relying on various physical science principles to replicate electro-chemical properties without resorting to current subscriptions to standard metals[ix].

# METHODS

Quite possibly, the most important characteristic associated with the production of photovoltaic paint is the actual schematics designed to construct the proper, stoichiometric-like blend. The beckoning of the various disciplines employed a number of materials, ultimately contributing to the final photovoltaic product.

To produce the paint, any water-soluble paint is acceptable, ex-post facto the conclusion. As for the experimentation, "American Tradition" Ultra Premium Valspar-brand crystalline-silica solution painting fluid was employed. The volume calculation was completed utilizing the ionic solution constituents, specifically, pure-grade Armor Table Salt, verifiable via "Technika Sper" scientific salt refractometer, 300006 (calibration-certification), as well as the testing unit Coralife Scientific-Grade Salt Mix, complete with calcium and magnesium fortified for high ionization.

The average inductance of all trials consists of levels approaching normal solar machinery. An alternative is to solve for ratios of wattage as a function of power, which is usually graphically presented as:

$$\frac{p(-W_p)}{p(W_n)} = \exp(q[V_{bi} \pm V / kT])$$

Which can be reduced to:

$$\frac{p(W_n)}{P_n} = e^{qV/kT}$$

Or expressed without the base-power factor

$$\frac{n(-W_p)}{n_p} = e^{qV/kT}$$

To ensure the maximum solubility without tainted or skewed results, pure distilled water was extensively utilized, with laboratory-grade, Model G-1, stainless steel heating element water distiller, matted as 99.9% effective. Testing revealed the impurity levels of bacteria, viruses, cysts, chemicals, heavy metals, organic minerals, and inorganic minerals at 0 ppm, truly demonstrating the potency of the laboratory-grade distiller.

Utilizing simple Flynn-grade measurement tools and flasks, the volume calculation process augmented and verified the exact contents of the ensuing calculations. Solute concentrations are generally reported either per unit mass or per unit volume of solvent. In this experiment's case, the solute concentration will be as measured per unit volume of solvent for the initial purveyance of the experimentation.

The layer of paint with the relatively lower concentration of ions, wired to the cathode, upholds a barrier region (known as the depletion zone) nestled between the said layer and the p-type replication. The insulation stage—portraying the properties of a rudimentary fuse—can be effectively omitted and the paints can come directly in contact with each other, catalyzing the flow of electrons and the establishment of a current. (Insulation allows effective omission due to the thermal equilibrium of the paints existing in said state prior to the experimentation—constants, such as temperature, humidity, and air pressure, are held throughout progression).

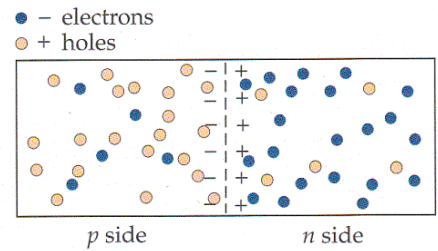

*Fig. 2: Rough visualization of conduction layer production (overlap)*

This demonstrates the primary objective when producing the paint solution prior to determining the volume calculations of the entire, reconstructed system. The photovoltaic paint theory relies on this sole information in determining the validity of the voltage generation. Solute concentrations are generally reported either per unit mass or per unit volume of solvent[x].

In this experiment's case, the solute concentration will be as measured per unit volume of solvent for the initial purveyance of the experimentation. Percentages by mass, however, serve as a check to the work, and the calculations are available to computerized-process via a Prostate Cell Program designed for biomedical instigations.

A common mass solute—mass solvent unit—is percent by mass, or $w/w\%$. Calculating concentration in this percentage is convenient because information about the solute's chemical nature is irrelevant to the concentration. One liter (L) of water possesses a mass of approximately one kilogram (kg). Example: To create a 10% ionic solution, 100g of Sodium Chloride (common table salt) to exactly 900mL of hydrogen dioxide (water).

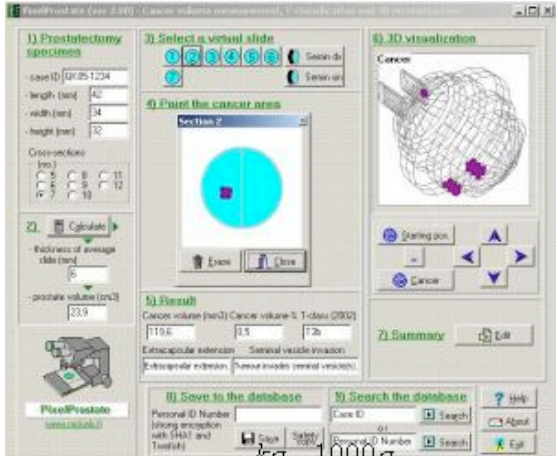

$$10\% \Rightarrow 0.1 \times 1L \times \frac{kg}{L} \times \frac{1000g}{kg} = 100g$$

*Fig. 3: From top to bottom—Prostate cell Program for biological simulation for volume calculation checks and the percent by mass equation.*

The paint production process requires meticulous attention to the proper ratio of ionic solution and water quantity, as well as actual ratios of the differing ionic solutions between the replicate layers of paint themselves.

The production process, as detailed above, is preserved after vehement stirring in special-coating containment units as sampling. The sampling space, however, would be taken from a home improvement store, where boards of all varying properties are consumable merchandise. The highest-quality board was selected to best foster the electrical current generation, and the store brand, 48"x24" sealant-primed wooden plasterboard with thermo-conductive resistance perfected the fit.

The 10% ionic solution batch was used the most often, as the base quantity for comparative electrical energy was construed to maximize the difference, for experimental purposes. The initial layer was applied to a single 8"x12" hand cut piece, and the Duraplex, high-quality chemistry measurement units were set aside as the paint source was currently unnecessary. Prior to the paint application, the Ballmer-brand steel apparatus with steel-netting and straightedge platform was the base of operations for the paint production facilitation and confined within a climate-controlled laboratory setting.

The initial application is designed to replicate the "doped" n-type conducting band of the conventional photovoltaic apparatus for solar panels. Therefore, for complete circuitry, the anode lead from the RadioShack-brand alligator-clamp voltmeter adapter unit with <1% error was attached to the Nicolet Biomedical EMG Surface Electrode (1.0 meter), in contact with the n-type band. The varying concentration greater than 10% (as 10% as the p-type band would eliminate the gradient for electron conduction flow), fortified structurally with a Bracken Sterilities Unit (chromium attachment), and complete with the Gauze Biomedical Tape (Kaiser Permanente), with security-latched porous ridges for flow capability, would also utilize the "Whizz" Roller System (high-density foam paint roller engineered for ultra-smooth surfaces) for application.

The second layer would need severe overlap with the initial paint layer to replicate the depletion zone that is so critical to the success of the current, conventional photovoltaic machinery. This depletion zone, however, relies on a simple overlap with broadening increasing the gradient and, therefore, increasing the voltage supplied. However, simply limiting the depletion zone to the majority of the board supplied addresses the scientific inquiry at hand: is photovoltaic paint possible? The p-type replication layer is connected with the cathode electrode, and also situated with the "alligator-clamp" voltmeter adapter unit before a complete circuit is created. The complete circuit is rudimentarily diagrammed.

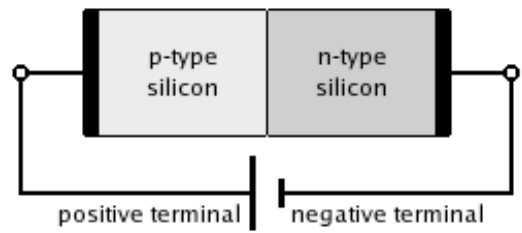

*Fig. 5: Essential p-n junction diode principle skeleton achieved via experimentation with ionic solution, electro-chemical gradients*

The complete circuit essentially resembles the above diagram at the most basic of levels, as the p- and n-type silicone adjacency represents the depletion zone the overlap band effectively replicates. The two lead "alligator-clamps" attached to the respective cathode and anode electrodes dependent entirely on the concentration of ionic solution within the paint solution.

The following concentrations were used and replicated on each board: 20%, 30%, 40%, and 50%. The results were recorded from the 22-Range, Auto-Ranging Pocket Multi-Meter RadioShack Voltmeter and electronically transcribed on an Excel Worksheet on a Dell Inspiron 9800 Personal Laptop Computer with graphing capabilities included. The entire operation of experimentation was conducted on the 5 m square foam insulation padding, which was designed as an experimental set-up area. The graphs, which will be shown in later sections, were composed via Shareware- Connection 3D Multi-Series professional graphing program with Microsoft-certification.

As an ending, all measurements were conducted via Flynn Scientific Weight Measurement for precision with electronic LED (+/- 1.0E-7 g error) and the DPD-1850/3030/6015 Laboratory Grade Triple Output Dual Tracking DC Regulating Power Supply was operated only as a source for reserve power. However, no time during my own experimentation was this device utilized.

## RESULTS

Prior to any mixture of ionic solutions into the conventional crystalline-silica paint, lone samples of unadulterated paint were subjected to sunlight and subsequently observed[xi]. With an effective difference of 0% ionic solution mixture due to 0%-0% duality, the solar energy collected by the painted panels was an expected negligible quantity and for the next 50 results, the current and voltage were at an unsurprising zero. As expected, the resistance, by Ohm's law, cannot be defined, as the value is 0/0. The purpose of this particular control grouping lays in the ideal of validating the voltage generation and current production via non-intrinsic means[xii].

Electrical energy, stemming form the electron migratory patterns, thereby, should, in theory, originate from the difference in the ionic solution concentrations per experimental units, discounting any possibility of crystalline-silica natural processes of electrical generation/production.

For a disparity of 10%, that is 10 and 20 per cent salinity, the average volts generated was 203.491 V, with a Standard Deviation: 11.319 V and a Population Variance: 128.120 V, along with a Magnitude of Current (In Amperes) Mean: 1.498 a Standard Deviation: 0.029 A and Population Variance: 0.008 A. The Magnitude of Resistance (In Ohms) possessed a Mean: 159.493, a Standard Deviation: 12.913, and a Population Variance: 166.746.

With 10% Salinity v. 30% Salinity, the Voltage Generation (In Volts) possessed an Average: 0.5217, a Standard Deviation: 0.5716 and a Population Variance: 0.0033 with a Magnitude of Current (In Amperes) Mean: 2.546, a Standard Deviation: 0.3731 and a Population Variance: 0.1392. The Magnitude of Resistance (In Ohms) possessed a Mean: 206.033, a Standard Deviation: 18.638, and a Population Variance: 347.387.

The concentration of 10% Salinity v. 40% Salinity, the Voltage Generation (In Volts) had an Average: 0.807, a Standard Deviation: 0.0287, and a Population Variance: 0.0008. The Magnitude of Current (In Amperes) was with a Mean: 3.498, a Standard Deviation: 0.193, and a Population Variance: .0372. The Magnitude of Resistance (In Ohms) possessed a Mean: 247.083, a Standard Deviation: 11.117, and a Population Variance: 123.588.

Combining this data, the electron flow can be calculated using

$$I_{ph} = qA \int_0^W G(x)dx$$

And the generation rate can be solved for,

$$G(x) = \frac{I_0}{\hbar \omega} \alpha \exp(-\alpha x)$$

Finally, the rate of wattage can be calculated,

$$I_{ph} = qA \frac{I_0}{\hbar \omega}[1 - \exp(-\alpha W)]$$

Which we use to solve for the Inductance, as a function of voltage:

$$I(V) \propto (e^{eV/kT} - 1)$$

By reducing many of these equations and solving for common variables (a task as easy as it sounds),

the biases of the photovoltaic apparatus can be graphically depicted, and the asymptotes are revealed, indicating infinite resistance in cases of breakdown voltage and forward bias compoundability:

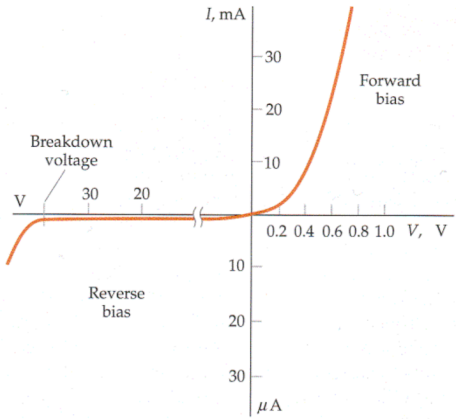

Using the defined technique of forward-bias calculation for each trial is notoriously inefficient, since each salinity concentration is independent of the inductance isolation pattern, a derivative of the difference between solubility.

The final concentrations were with 10% Salinity v. 50% Salinity. The tests revealed a Voltage Generation (In Volts) with an Average: 1.219m a Standard Deviation: 0.169, and a Population Variance: 0.029[xiii]. Conversely, the Magnitude of Current (In Amperes) was with a Mean: 4.329, a Standard Deviation: 0.075, and a Population Variance: 0.006. The Magnitude of Resistance (In Ohms) possessed a Mean: 339.762m Standard Deviation: 18.976, and a Population Variance: 360.089. The data is discussed in the subsequent section.

## PRACTICALITY

Photovoltaic paint derives its functionality from its application, both in vast scale and scope; that is, because paint can be applied to virtually any surface (sealant-pending)[xiv], solar power is capable from virtually any location on the surface of the Earth (sun-pending). The universality of the theoretical basis is unimaginably vast, as the application does not limit itself to specialized locations where hulking, aluminum frames might otherwise require planning-specific constructions. The ability to plaster a surface with the photovoltaic paint implicates an epoch of novel pathways to generate energy for an exponentially rising population continuously requiring larger and vaster quantities of energy over time from a limited pool[xv]. The experiments reveal the very tangible possibility of realizing the limitless accessible energy and usage from both the practical standpoint for the mechanical merits, and the aesthetical perspective for artistic maintenance. Of course, the relatively small voltage harvest ushers further research into improving the electrical conversion capabilities[xvi] of the photovoltaic paint, and the conduction layers of the ionic solution may increase with chemically-perfected techniques geared towards maximizing the current flow. As far as the function of temperature is concerned, an upper-limit exists for the derived eV per temperature unit, and is outlined below.

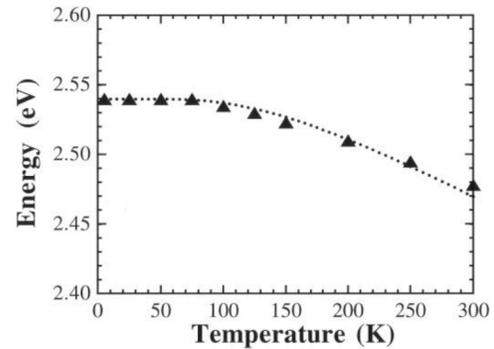

An array of applications flood the experimental market, as the idea of providing sufficient energy in electrolysis in hydrogen fuel-cell automobiles, or generating sufficient voltage to run entire buildings, enters the realm of possibility[xvii]. The perfection in maximizing the voltage production of the photovoltaic paint establishes new ground for cost-effective and unlimited applicative possibilities in energy-saving endeavors and advances. As the experimentation may show, the application of several differing layers of semi-conducting polymer paint to the exterior of buildings, motor vehicles, and anything else exposed to sunlight would, in all probability, reap and yield enormous benefits in energy capture, especially in equatorial and tropical locales. Consequentially, the photovoltaic paint would alleviate the visual pollution of currently existed solar arrays, so as long as the arrays themselves are plastered with the material as well. Future experimentation, no doubt, offers enticing advancements in paint composition-sodium-chloride-ratio examinations, as well as the efficiency-catalyzing production methods stemming from meticulous re-examination of key consistency errors that may have, in fact, materialized during the course of testing. The possibilities of both reconstituting the current procedural bearings, as well as the actual ionic solution offer limitless variations in constructing optimum values of energy yield.

Lacking in my project was the continuation of further values of ionic solution, as my experimentation seized at the expected maximum values of efficiency in comparison to conventional PV crystalline-silica panels. Further experimentation includes the factoring in of weather variations that may skew results, such as unsuitable extremes in temperature and precipitation.

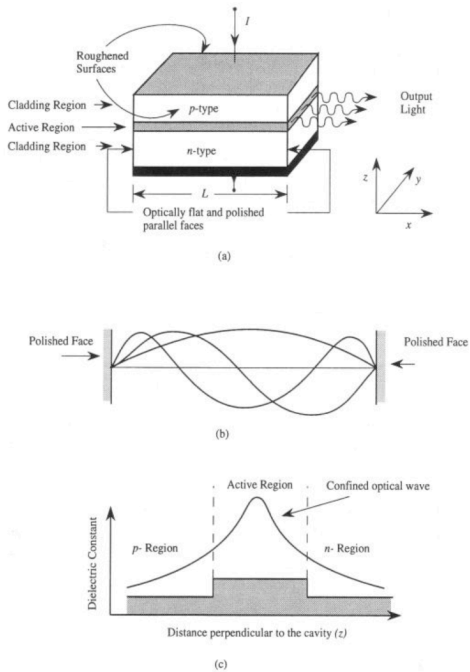

In fact, the mere mentioning of the alternative energy panels beckons cross-collaboration in terms of contemporary practices. Crystalline semiconductor panels—traditional photovoltaic arrays—can be augmented by the addition of a photovoltaic paint coating. Wind-farms, characterized by acres of propellers harnessing the wind energy, can, too, feel the coating of photovoltaic paint in order to generate natural means of electrical energy as a multi-faceted source of alternative energy. One very important area of further study is the lifespan of these solar materials, as opposed to many durable solar panels with a virtually 30 year lifespan to even lifetime use depending on the durability (the thin layer of crystalline) construction and sustainability. Factual evidence exists voicing the concern over the photovoltaic delicacy of this aforementioned crystalline layer, and further studies incorporating my work into their experimentation. Solar panels, when heralded by increasing demand, allow, by rudimentary economic models, a price decrease with an increase in supply. The cost per unit, therefore, becomes considerably less than current standards of prices. As aforementioned, the weathering extremes, concentration of salinity, and the utilization of differing ionic solutions for maximum voltage output were based on intrinsic properties of the traditional, crystalline silica paint. This is due in large part to the

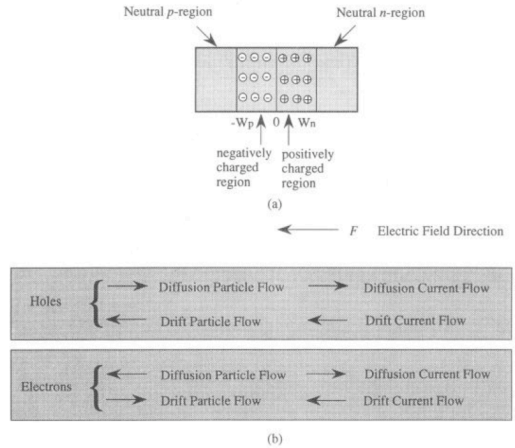

The possibilities for application, however, remain virtually limitless. The implications of photovoltaic paint are vast, but no such hasty coronation may continue prior to the analysis of the data. However, the fact that actual voltage production occurred, unhindered by excessive resistance, although, suffice to say, sizable quantities of resistance were present[xviii], merits achievement in principle.

The indication of the control group reveals absolutely no voltage or current generation, and, as equated by Ohm's Law, no resistance can be calculated as the unit, ohm, depends on nonzero integers for real values. The purpose of the control grouping as a comparative tool to variable experimentation is to validate the hypothesis. In other words, if any electrical generation does, in fact, occur, the electron movement is not due to the replication of the thermoelectric effect[xix] and the natural crystalline-silica formation of the paint fluid itself. However, as aforementioned, no voltage or current was actually generated, implying that if any electrical energy were to surface, the cause would originate from the ionic solution itself and not the painting solution. With a 10% difference in salinity, the photovoltaic paint generates about one-fifth of a volt, with 20%, the difference amounts to approximately one-half a volt[xx].

A 30% salinity difference reveals just about three-quarters of a volt, whereas the maximum tested potential, a 40% difference, reveals 1.25 volts per square foot. The average resistance, revealed by the Statistical Analysis elsewhere, indicates a very large quantity of ohms impeding the current magnitude. For the first experimental group, an average of 150 ohms generated the magnitude of resistance, while the 10% v. 20% congregation revealed the maximum cap at 200 ohms. The third and fourth order experimental groups yielded, on average, 2.5 and 3.5 mA, respectively; finally, the 50% P-type ionic solution generated around 4.5 mA of electrical energy. The value of the current causes the extremely large quantity of resistance, as the values, measured in milliamps, are extremely miniscule in relation to the voltage produced, as Ohm's Law dictates. For a 10% salinity difference, the average milliamp value strays just over 1.5, and 20% delivers one more entire milliamp.[xxi]

With a 30% ionic concentration difference, an additional milliamp is produced, with the final ionic solution generating 4.5 milliamps of current. The distribution spread, as computed by the population standard deviations and variances, indicate a relatively set grouping of values, revealing the relatively proximal approximation of values hovering in the same range. The variance and the closely related standard deviation are measures of, as aforementioned[xxii], how spread out a distribution is. In other words, they gauge the variability of certain results. The variance is computed as the average squared deviation from each experimental unit from the mean itself. The standard deviation formula is quite simple: the square root of the variance. It prides itself as the most common measure of spread.

An important attribute of the standard deviation as a measure of spread is that if the mean and standard deviation of a normal distribution are known, it is, in fact, quite possible, compute the percentile rank associated with any given score, known as a z-score. The ionic salinity concentrations are directly proportional to the electrical voltage generated, as well as the magnitude of the quantity of amperes. Consequentially, the magnitude of the resistance, measured in ohms, rises significantly, and, as more voltage and current is produced, the resistance increases dramatically, eventually negating any production of further electrical energy. The projected values augment the reasoning for limiting the experimental units to the maximum salinity difference to approximately 40%.

The initial testing was completed on a control board of unfinished, un-primed modular wood-panels, where the voltage generation may exhibit inhibited function, and must be administered in a neutral environment.[xxiii] By ensuring multi-surface applicability, the photovoltaic paint can enjoy universal application, and then climate studies can be performed. If the lifespan can be determined from the context to which it is deployed, then exhaustive testing is required to determine the chemical degradation of weathering[xxiv], including excessive photonic flux and a number of water-based precipitates. However, the main concern regarding the photovoltaic paint is the instantaneity of the voltage charge, and the issue of the possibility of lengthening the duration.

Photovoltaic Paint is a novel vector for delivering highly affordable alternative energy to the populace with a complex juxtaposition of simple parts.

---

xix A
xx
xxi
\

CONTACT: Arjan Singh Puniani; 559-392-1572;
e-mail correspondence:
arjan.puniani@berkeley.edu